\documentclass[referee]{kluwer}    
\usepackage{graphicx}
\newdisplay{guess}{Conjecture}
\begin{document}

\newcommand {\Tef} {T$_{\rm eff}$~}
\newcommand {\la} {$\lambda$~}
\newcommand {\lala} {$\lambda$$\lambda$~}
\newcommand {\vt} {$V_t$~}
\newcommand {\nli} {log N(Li)~}
\newcommand {\dli} {$\Delta$log N(Li)~}
\newcommand {\kap} {$\kappa$~}
\newcommand {\CTWO} {C$_{\rm 2}$~}
\newcommand {\CCr} {$^{12}$C/$^{13}$C~}
\begin{article}
\begin{opening}
\title{\bf
Lithium abundances in the atmospheres of SLR 
C--giants WZ Cas and WX Cyg from resonance and subordinate Li I 
lines} 

\author{Larisa A. \surname{Yakovina}}
\author{Yakiv V.  \surname{Pavlenko}}
\institute{Main Astronomical Observatory of
NASU, Golosiiv woods,
 Kyiv-127, 03680, Ukraine\thanks{e-mail: 
 yakovina@mao.kiev.ua, yp@mao.kiev.ua}}
\author{Carlos    \surname{Abia}}
\institute{Dpt. F\'\i sica Te\'orica y del Cosmos, Universidad 
de Granada, 18071 Granada, Spain\thanks{e-mail: cabia@ugr.es}}
 
\runningauthor{L.A.Yakovina et al.}\runningtitle{
Lithium abundances in atmospheres of SLR C--giants ...}

\begin{abstract}

Lithium abundances in the atmospheres of the super Li-rich 
C-giants WZ Cas and WX Cyg are derived by the spectral synthesis 
technique {\bf using} the Li I resonance line at \la670.8 nm and 
{\bf three subordinate} lines at \lala 812.6, 610.4 and 497.2 nm. 
The differences between the Li abundances derived from the \la670.8 
nm line and the \lala 497.2, 812.6 nm lines do not exceed $\pm$ 
0.5 dex. The lithium line at \la610.4 nm provides typically lower 
abundances than the resonance line (by $\approx$ 1 dex). The mean 
LTE and NLTE Li abundances from three Li I lines (excluding 
\la610.4 nm) are 4.7, 4.9 for WZ Cas, and  4.6, 4.8 for WX Cyg, 
respectively.

\end{abstract}

\end{opening}

\section{Introduction}

{\bf The knowledge of lithium}
abundances in stellar atmospheres {\bf is 
important because it provides} 
information about physical processes 
in stars and stellar evolution. {\bf Some asymptotic giant branch (AGB) 
carbon stars (C/O$>$1, by number) are observed to be lithium rich with 
respect to the solar value of \nli = 1.2 (in scale 
log N(H)=12). WZ Cas and WX Cyg are well known super Li rich (SLR, 
\nli $\geq$ 4) giant carbon stars.} 
{\bf AGB SLR} stars{\bf,} with {\bf their 
strong} mass loss{\bf,} could be the main suppliers of lithium 
in {\bf the} Galaxy (cf. Abia et al. 1993, Wallerstein \& Knapp 1998). 
{\bf Furthermore, the determination of Li abundances in the atmospheres 
of these stars has also cosmological interest.} 

Currently, the overwhelming majority of determinations of the 
lithium abundances {\bf in 
carbon} stars are based on the 
analysis of the resonance Li I line \la670.8 {\bf nm only}. 
However, spectra of {\bf Li rich} 
stars contain
{\bf several Li lines that are possible probes of the Li abundance.}
{\bf In
Yakovina} \& Pavlenko (2001), six subordinate Li I lines in the 
region \lala 400-820 nm were analysed.
{\bf The present paper examines which of them can be used as Li 
abundance indicators in carbon stars, and the ranges of Li abundances 
over which they work.} This {\bf paper}
is an extension of {\bf that}
by Abia et al. (1999), in which the formation of the Li I lines 
in atmospheres of SLR AGB stars 
{\bf was} analysed.

We use {\bf echelle-}spectra of 
WZ Cas (C9.2J) and WX Cyg 
(C8.2eJ) obtained in 1997--1999 with the 4.2 m WHT (Roque de los 
Muchachos observatory) and 2.2 m telescope (Calar Alto 
observatory; see Abia et al. (1999) for more details). {\bf The 
spectral resolution was $\lambda$/$\Delta\lambda$ $\approx$ 
50000 and 35000, respectively.}

\section{Lithium abundance indicators.}

The strong subordinate Li I lines \lala 812.6, 610.4, 497.2, 
460.3, 427.3 and 413.3 nm are formed by transitions from the 
level 2p. 
{\bf However, only the first three turn out to be useful for
measuring Li abundances: the \la460.3 nm line is severely blended  
with a Fe I line at \la460.3 nm, and the \la427.3 nm line with a Cr I
line at \la427.5 nm. In addition, the Li lines at \la427.3 nm 
and \la413.3 nm usually cannot be observed because of the strong violet 
depression in the spectra of carbon stars at these wavelengths.}

{\bf Our analysis below shows that the \la497.2 nm line is a 
good} 
indicator of the lithium abundance in the 
range  3$<$log N(Li)$<$5, 
{\bf lying} in a {\bf spectral} region 
{\bf only moderately obscured by CN and \CTWO absorptions.} 
The lithium lines at \la812.6 nm and \la610.4 nm are 
saturated for \nli$>$3 
{\bf but} they are of high sensitivity {\bf on Li 
abundance} in the range 2$<$log N(Li)$<$3 (Abia et al. 1999). 
{\bf The only line strong enough to 
measure lower Li abundances is the \la670.8 nm resonance line.}

In this paper we determine and compare the {\bf LTE and NLTE} 
lithium abundances derived in  WZ Cas and WX Cyg from theoretical 
fits to the resonance Li I line \la670.8 nm and to the 
subordinate Li I lines \lala 497.2, 610.4 and 812.6 nm. {\bf As 
was recently shown by Abia et al.(1999), it is essential to 
perform a NLTE analysis when deriving Li abundances in SLR carbon stars 
due to importance of non-equilibrium processes in the resonance 
\la670.8 nm Li line.}

\section{Synthetic spectra computations}

Synthetic spectra were computed in the framework of LTE approach 
using the program WITA6 (Pavlenko 2000). Some specific continuum 
opacity sources in carbon-rich atmospheres (bound-free absorption 
of C I, C$^-$, O I and N I) were taken into account. In some 
cases we implement an additional continuum absorption to account 
for some not yet identified opacity sources
{\bf (increasing the continuum opacity} by {\bf a} factor \kap. 
{\bf We obtain NLTE 
Li abundances using LTE and NLTE curves of growth as computed in
Abia et al. (1999).}

\begin {table}
\caption {Parameters of the model atmospheres for WZ Cas and
WX Cyg}

\begin {tabular} {ccccccccccc}
\noalign {\smallskip}
\hline
\noalign {\smallskip}
 Star&\Tef/log g/[$\mu$] & C/N/O$^*$ \\
\noalign {\smallskip}
\hline
\noalign {\smallskip}
WZ Cas & 3000/0./0. & 8.923/7.99/8.92 \\
\noalign {\smallskip}

WX Cyg & 3000/0./0. &  8.93/7.99/8.92 \\

\noalign {\smallskip}
\noalign {\smallskip}
\hline
\noalign {\smallskip}
\end {tabular}

$^*$ - in scale log N(H) =12.0 \\

\end {table}

The model atmospheres{\bf, N and O abundances} 
are {\bf taken} from 
{\bf the} grid of Eriksson et al. (1984) (see Table 1). 
Carbon abundances were 
{\bf slightly adjusted
(by less than 0.004 dex) using the} fits to observed spectra
. Computations were carried out using a 
microturbulence velocity \vt=2.5 km/s and a isotopic ratio \CCr=5 
(cf. Abia et al. 1999) for both stars.

\section{Atomic and molecular line lists}

Different sources of atomic and molecular lines were used:

-- VALD (Kupka et al. 1999) and Kurucz's (1993-1994) databases;

-- The line lists used by Abia et al. (1999);

-- The line list kindly provided by T.Kipper which consists of an 
atomic line list from R.Bell (private communication) and 
molecular line lists from D.R.Alexander (private communication).

Atomic {\bf line}
data were verified {\bf by comparing} 
synthetic spectra computed 
with the solar model atmosphere of Kurucz (1993-1994) to the 
observed spectrum of the Sun (Kurucz et al. 1984). The fits 
obtained to the solar spectrum in all regions studied were rather 
good (Yakovina \& Pavlenko 2002). 

The main electronic systems of diatomic carbon-containing 
molecules accounted for  in our synthetic spectra computations 
are listed in Table 2. The last column of Table 2 contains the  
dissociation potentials {\bf used.} 

\begin{table}
\caption{Molecular systems of diatomic molecules considered in
synthetic spectra calculations}
\begin{tabular}{ccccccc}
\noalign{\smallskip}
\hline
\noalign {\smallskip}
\noalign {\smallskip}
 Molecule & Transition & Name of system & D$_0$ (eV) \\

\noalign{\smallskip}
\hline
\noalign{\smallskip}
$^{12}$C$^{12}$C, $^{12}$C$^{13}$C, $^{13}$C$^{13}$C& d$^3\Pi_g$ - a$^3\Pi_u$  &  Swan   & 6.156 \\
$^{12}$C$^{12}$C, $^{12}$C$^{13}$C, $^{13}$C$^{13}$C & A$^1\Pi_u$ - X$^1\Sigma^{+}_g$ &  Phillips & 6.156 \\
$^{12}$C$^{14}$N, $^{13}$C$^{14}$N  & B$^2\Sigma^+$ - X$^2\Sigma^+$  &  Violet  & 7.89 \\
$^{12}$C$^{14}$N, $^{13}$C$^{14}$N  & A$^2\Pi$ - X$^2\Sigma^+$ &  Red  & 7.89 \\
$^{12}$CH, $^{13}$CH  & A$^2\Delta$ - X$^2\Pi$   &     &  3.47    \\
$^{12}$CH, $^{13}$CH  & B$^2\Sigma$ - X$^2\Pi$   &     &  3.47    \\
\noalign{\smallskip}
\hline
\end{tabular}
\end{table}

We analysed molecular line lists qualitatively using spectra of 
molecular electronic systems computed with line lists from 
different sources. The best versions were chosen from comparison 
of computed and observed spectra of WZ Cas and WX Cyg. Our main 
conclusions about molecular line lists are:

-- The CN line lists of Kurucz (1993-1994) and Abia et al. (1999) 
agree well.

-- The line lists of \CTWO Swan system of these authors provide 
theoretical spectra that agree well in structure but disagree in 
intensity.

-- In {\bf some wavelength intervals, near the Li lines,} 
the line lists by Abia et al. (1999) provide 
better fits to the observed spectra 
than the lists by Kurucz (1993-1994). On the other hand, the 
Kurucz line lists are more complete and uniform
{\bf: that is, it supplies the better fit to observed spectra 
in wider spectral ranges.} 

We mainly used the Kurucz's (1993-1994) molecular line lists from 
CD-ROM No.18 because with a more complete line list we are able 
to {\bf locate} more confidently 
the pseudo-continuum level.

\section{Results}

Fits to the spectrum of WZ Cas in four regions containing Li I 
lines are 
{\bf shown} in Fig. 1. Observed and synthetic spectra agree 
well enough, but there are still some discrepancies. We believe 
that they are mainly due to the lack of precision and 
completeness of {\bf the} molecular line lists used.

\begin {figure}
\includegraphics [width=12cm, height=19cm, angle=00]{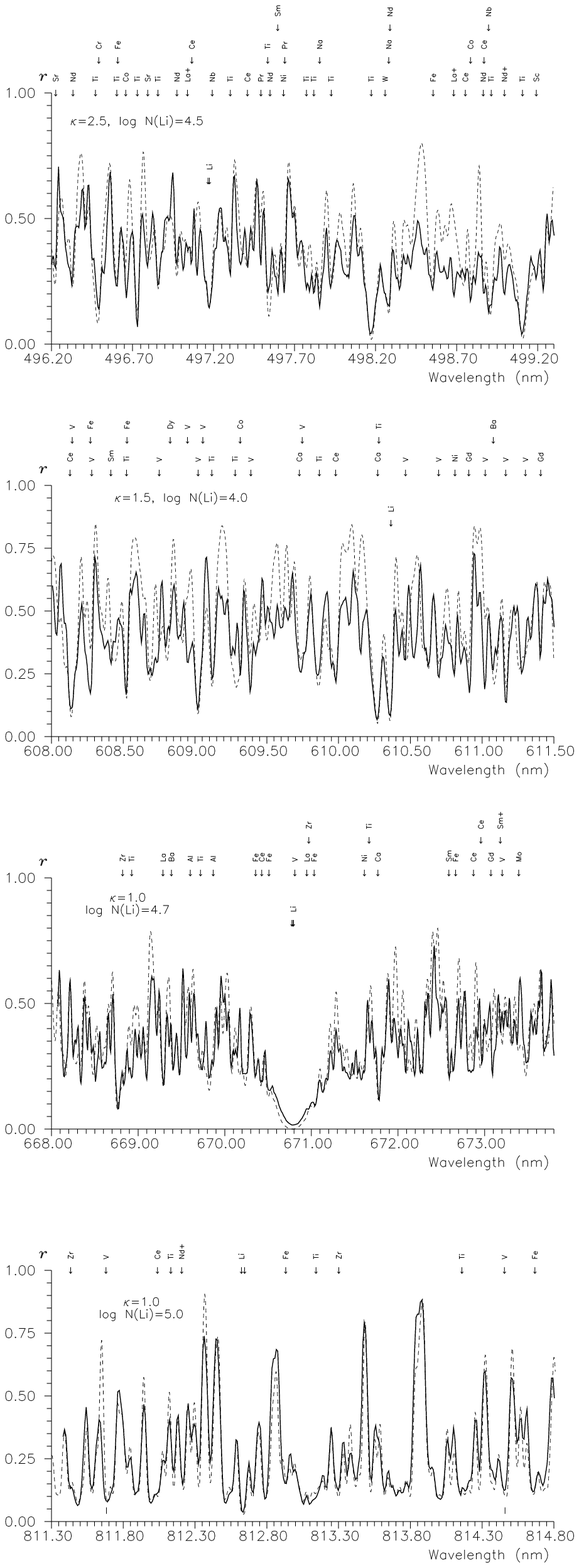} 
\caption{Fits
to observed spectra WZ Cas (solid line). The \nli and \kap are
specified for synthetic spectra (dashed lines). \kap=1 means
absence of additional continuous absorption. }

\end {figure}

The {\bf effect}
of blends containing Li I lines on \nli and the 
fits to the observed blends are shown in Fig. 2. The LTE and NLTE 
Li abundances obtained and the comparison 
{\bf with} the results by Abia 
et al. (1999) and Abia et al. (1991) are given in Table 3. {\bf Note
that} for WZ Cas we specified the range of uncertainty of the Li 
abundance derived from the Li I \la670.8 nm line, log N(Li)= 
4.7--5.0. In this case \nli was obtained by fitting the blended 
wings of the resonance Li I line in a region of about $\sim$ 2 
nm. This makes our fitting rather ambiguous. We choose ``the 
best'' fit to be \nli= 4.7.
 
\begin {figure}
\includegraphics [width=11cm, height=9cm, angle=00]{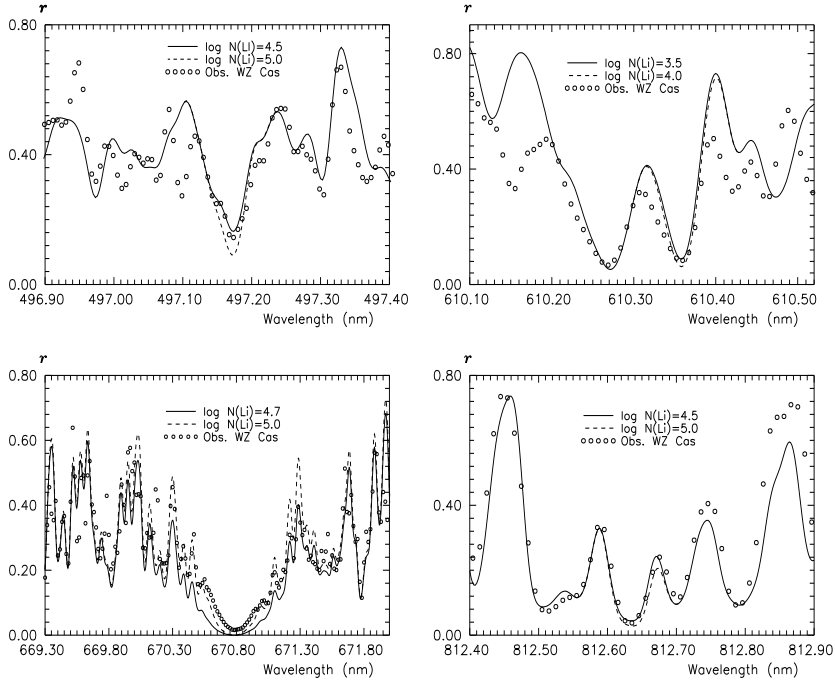} 
\caption [] {Sensitivity of blends containing Li I lines on
the Li abundance. Observed spectra are shown by circles}
\end {figure}

Table 3 shows the results obtained for different  molecular line 
lists taken from the Kurucz (1993-1994), Abia et al. (1999) and 
Kipper data. Differences in the \CTWO line lists change a little 
the carbon abundances, 
{\bf but the effect in \nli is lower than 0.1 dex.}

\begin{table}
\caption {LTE and NLTE lithium abundances in the atmospheres of
WZ Cas and WX Cyg.} 
\begin{tabular}{ccccccccccc}
\noalign{\smallskip}
\noalign{\smallskip}
\hline
\noalign{\smallskip}
\noalign{\smallskip}
 Star & Li I line (nm) & LTE & $\Delta_{LTE}$ & NLTE &
 $\Delta_{NLTE}$ & LTE APL & LTE ABIR \\
\noalign{\smallskip}
\hline
\noalign{\smallskip}
\noalign{\smallskip}
 WZ Cas &  497.2 & 4.7 &  0.0 & 4.8 & -0.2 &  -  &  -  \\ 
        &  610.4 & 3.7 & -1.0 & 3.8 & -1.2 & 3.1 &  -  \\ 
        &  670.8 & 4.7 &   -  & 5.0 &   -  & 4.5 & 5.0 \\ 
        &  812.6 & 4.7 &  0.0 & 4.8 & -0.2 & 4.6 &  -  \\ 
\noalign{\smallskip} 
\noalign{\smallskip}
 WX Cyg &  497.2 & 4.3 & -0.2 & 4.4 & -0.4 &  -  &  -  \\ 
        &  610.4 & 4.0 & -0.5 & 4.1 & -0.7 & 3.2 &  -  \\ 
        &  670.8 & 4.5 &   -  & 4.8 &   -  & 3.7 & 4.7 \\ 
        &  812.6 & 5.0 & +0.5 & 5.1 & +0.3 & 4.2 &  -  \\ 
\noalign{\smallskip} 
\noalign{\smallskip} 
\hline 
\noalign{\smallskip} 
\noalign{\smallskip} 
\end{tabular}

 $\Delta_{LTE}$, $\Delta_{NLTE}$ -- LTE and NLTE (log 
 N(Li)$_{\lambda}$ -- log N(Li)$_{670.8}$),  \\
   $\lambda$ - the wavelength of current subordinate Li line; \\
   APL - Abia et al. 1999; \\
  ABIR - Abia et al. 1991. \\
   
\end{table}

As it can be seem in Table 3, LTE and NLTE lithium abundances 
determined from subordinate Li I lines do not differ 
systematically from the estimations using the resonance line:
 
-- The \nli values derived from \la497.2 nm line are equal or a 
bit lower than estimations from \la670.8 nm line (\dli $\leq$ 
0.4).

-- Lithium abundances from \la610.4 nm line are essentially (by 
$\approx$ 1 dex) lower than the ones from the resonance line. 

-- The estimations from the line \la812.6 nm are approximately 
equal or higher than those from \la670.8 nm line. NLTE 
corrections improve the agreement between the lithium abundance 
estimations from the Li I \la812.6 nm and {\bf the} resonance line. 

We note that the derivation of the lithium abundance in the 
atmospheres of SLR carbon giants from saturated lines Li I \lala 
812.6 and 610.4 nm are less {\bf reliable}
in comparison with the \nli 
values from the unsaturated line Li I \la497.2 nm. Considering 
the LTE results in this work, Abia et al. (1999) for 3 stars and 
Plez et al. (1993) for 6 stars, we find that  log 
N(Li)$_{610.4}$--log N(Li)$_{670.8}$ and log N(Li)$_{812.6}$--log 
N(Li)$_{670.8}$ are in the ranges $-0.5$ to $-1.4$ and $0.0$ to 
$+0.7$, respectively.

Finally, for the lithium lines \lala 497.2, 670.8 and 812.6 nm we 
obtain the mean log N$_{LTE}$(Li) = 4.7 and 4.6 for WZ Cas and WX 
Cyg, respectively. The mean NLTE lithium abundances for the same 
lithium lines are 4.9 for WZ Cas and 4.8 for WX Cyg.

\section{Discussion}

Our results show that the differences in {\bf the}
lithium abundances {\bf estimated} from subordinate lines Li I 
\lala 497.2 and 812.6 nm 
and from {\bf the} resonance line Li I \la670.8 nm are of the same 
order {\bf: $\pm$ 0.5 dex. This is similar to} 
the possible errors in \nli due to 
uncertainties in the atmosphere parameters, model atmospheres and 
continuum level (see Denn et al. 1991, Abia et al. 1991, Plez et 
al. 1993, Abia et al. 1999).

The underestimations of {\bf the Li abundance from the $\lambda$ 
610.4 nm line might} 
be explained by the bad fits to the observed spectra in this region. 
On the other hand, as it was already mentioned, the saturated Li I 
line \la610.4 nm shows a rather weak dependence on log N(Li).

Finally, we conclude that the resonance \la670.8 nm and 
subordinate \la497.2 nm Li I lines are the best lithium abundance 
indicators in atmospheres of SLR carbon stars.
 
Our lithium abundances for WZ Cas and WX Cyg (Table 3) agree well 
with results of Abia et al. (1991) based on the resonance Li I 
line. {\bf Their Li estimations} 
are higher by 0.2--0.3 dex, probably due to 
the incompleteness of the line lists used. Lithium abundances 
{\bf by} Abia et al. (1999) agree well with our results for WZ Cas. 
For WX Cyg our 
estimations are higher by 0.7--0.8 dex. We believe 
that this difference is due to the different location of 
continuum level.

\acknowledgements

The authors thank the founders of a database VALD for qualitative 
and convenient service, Dr. T.Kipper for providing the line lists 
and Hugh Jones for careful proofread of the papers text. 
Financial support was provided by SRG of AAS. {\bf Authors thank
anonymous referee for very useful remarks.}

\end{article}
\end{document}